\documentclass[12pt,preprint]{aastex}

\def\bat{{{\it Swift}/BAT }}
\def\lat{{{\it Fermi}/LAT }}

\def\rx{{$R_X$ }}
\def\rbat{{$R_{X, BAT}$ }}

\received{}
\accepted{}
\slugcomment{Accepted, 8 Sept. 2011}
\shorttitle{Fermi View of Seyferts}
\shortauthors{Teng et al.}
\journalid{}{}
\articleid{}{}

\begin{document}

\title{\lat Observations of \bat Seyferts: on the Contribution of Radio-quiet AGN to the Extragalactic $\gamma$-ray Background}

\author{Stacy H. Teng \altaffilmark{1,2,3}, Richard F. Mushotzky \altaffilmark{2}, Rita M. Sambruna \altaffilmark{4}, David S. Davis \altaffilmark{3,5}, and Christopher S. Reynolds \altaffilmark{2}}

\altaffiltext{1}{Contacting author: stacyt@astro.umd.edu.}
\altaffiltext{2}{Department of Astronomy, University of Maryland,
  College Park, MD 20742, U.S.A.}
\altaffiltext{3}{CRESST and X-ray Astrophysics Laboratory, NASA/GSFC, Greenbelt, MD 20771, U.S.A.}
\altaffiltext{4}{Department of Physics and Astronomy, George Mason University, Fairfax, VA 22030, U.S.A.}
\altaffiltext{5}{Department of Physics, University of Maryland Baltimore County, Baltimore, MD 21250, U.S.A.}

\begin{abstract}

We present the analysis of 2.1 years of \lat data on 491 Seyfert galaxies detected by the \bat survey.  Only the two nearest objects, NGC~1068 and NGC~4945, which were identified in the {\it Fermi} First-year Catalog, are detected.  Using the \bat and radio 20~cm fluxes, we define a new radio-loudness parameter \rbat where radio loud objects have $\log R_{X, BAT} > - 4.7$.  Based on this parameter, only radio loud sources are detected by \lat.  An upper limit to the flux of the undetected sources is derived to be $\sim 2 \times 10^{-11}$~photons~cm$^{-2}$~s$^{-1}$, approximately seven times lower than the observed flux of NGC~1068.  Assuming a median redshift of 0.031, this implies an upper limit to the $\gamma$-ray (1--100~GeV) luminosity of $\lesssim 3 \times 10^{41}$~erg~s$^{-1}$.  
In addition, we identified 120 new \lat sources near the \bat Seyferts with significant \lat detections.  A majority of these objects do not have \bat counterparts, but their possible optical counterparts include blazars, FSRQs, and quasars.

\end{abstract}


\section{Introduction}
\label{sec:intro}

Recently, the {\it Fermi}/Large Area Telescope (LAT) team has released the first year catalog of LAT-detected active galactic nuclei (AGNs) with greater than 5-$\sigma$ significance \citep{firstyr}.  Along with many blazars and several radio galaxies, a few Seyfert galaxies (specifically NGC~1068 and NGC~4945), traditionally classified as radio-quiet sources, are listed as possible counterparts to the \lat sources.  \citet{nlsy1} also reported the \lat detection of $\gamma$-ray emission from the narrow-line Seyfert~1 galaxy PMN~J0948+0022.  The detection of type-1/2 AGN at GeV energies represents a major breakthrough for {\it Fermi}, as it points to a new class of $\gamma$-ray emitters.  Observationally, this is an exciting discovery as no Seyfert was previously detected with EGRET.  The canonical spectral energy distribution (SED) of a radio-quiet AGN has a turnover around several hundred keV \citep[e.g.,][]{dermer}; GeV $\gamma$-ray emission is usually not part of our standard view of these systems.  However, there are theoretical reasons to expect significant $\gamma$-ray emission from the active nucleus of Seyferts of all luminosities, and depending on their $\log N - \log S$, Seyferts could turn out to be non-negligible contributors to the extragalactic $\gamma$-ray background (EGB), contrary to current claims \citep{ajello}.  

From a theoretical point of view, $\gamma$-ray emission from the active nucleus in Seyferts is not unexpected. At sub-Eddington luminosities, it is quite likely that accretion proceeds via radiatively inefficient flows \citep[e.g., ADAF;][]{adaf}.  Here, electrons are heated to temperatures $T_e \sim 10^9$~K via interactions with protons of $T_p \sim 10^{12}$~K, and produce the observed radiation via bremsstrahlung and inverse Compton scattering of the disk optical-IR photons, producing a bump from IR to X-rays with hard X-ray continua.  A general feature of ADAF models is that they predict a second bump peaking around 10$^{23}$~Hz due to the decay of pions produced in the collisions of the hot proton gas \citep[e.g.,][]{mahadevan, oka}.  The exact shape and normalization of the $\gamma$-ray peak are a function of a number of parameters, including the black hole spin, since for larger values of the spin the disk extends deeper in the gravitational well with higher proton temperatures and hence higher X-to-$\gamma$-ray flux ratios.  For high spin values (a=0.95) and favorable values of the other parameters, the X-to-$\gamma$-ray flux ratio $\sim$~1.  The peak emission is near 100~MeV and decreases very sharply with energy in models without non-thermal protons.  Thus, by measuring the \lat flux and, possibly, the spectrum, it is possible to constrain several important quantities in the ADAF, e.g., the black hole spin and the gas composition.

Another possible scenario for $\gamma$-ray emission in radio-quiet sources may be due to coronal effects.  At higher luminosities ($\sim$~1\% Eddington or larger), the conventional view is that the X-rays are produced via inverse Comptonization of soft (UV/optical) photons from a radiatively-efficient accretion disk by electrons in a hot disk corona.  The corona is believed to be heated by magneto-hydrodynamic (MHD) waves or magnetic reconnection associated with MHD turbulence in the underlying disk \citep[e.g.,][]{miller}.  While the corona is normally treated as a thermal plasma with temperature 100--200~keV, the corona is likely to be a marginally collisionless plasma \citep{goodman} and this raises the possibility that non-thermal particle acceleration may accompany the magnetic heating.  The detection of a hard $\gamma$-ray tail beyond the ``thermal cutoff'' of the X-ray spectrum would be strong evidence for a population of non-thermal accelerated electrons within the corona.

Finally, the $\gamma$-ray emission may be produced by interactions between possible jets and the immediate environment of the active nucleus.  Some Seyferts with deep radio imaging do exhibit parsec-scale ejecta, similar to radio-loud sources \citep[e.g.,][]{nagar}.  An example is NGC~1068, a type-2 source with a parsec-scale radio jet interacting with a molecular cloud near the nucleus \citep{gallimore}.  While the jets are in general only mildly relativistic, and beaming is not strong enough to amplify non-thermal $\gamma$-ray emission, the jet-ISM interaction could in principle be responsible for the production of $\gamma$-rays via hardonic processes (proton-proton collisions), as observed in starbursts.  


The \bat sample of nearby Seyfert galaxies consists of non-blazars detected with \bat in the 14--195~keV energy range.  In the following section, we will discuss the \bat sample and our {\it Fermi} data analysis.  In \S~\ref{sec:disc}, we will discuss the results and their implications; a summary of our findings is presented in \S~\ref{sec:sum}.  Throughout the rest of this paper, we assume the cosmology H$_0 = 71$~km~s$^{-1}$~Mpc$^{-1}$, $\Omega_M = 0.27$, and $\Omega_\Lambda = 0.73$.  The {\it Fermi} flux will be measured in the 1--100~GeV energy band where the effective area of \lat is the highest.

\section{Sample Selection and {\it Fermi}/LAT  Data Analysis} 
\label{sec:obs}

\subsection{The \bat Sample of Seyferts}

The Burst Alert Telescope (BAT) on board {\it Swift} operates in the 14--195~keV energy range and with a field of view of $\sim$10\% of the sky.  BAT observes 60\% of the sky on average each day at the 20~mCrab level.  After 58 months of operation, the \bat catalog has identified 519 Seyfert galaxies and 108 beamed AGN sources in addition to many pulsars, X-ray binaries, and other classes of objects \citep{batcat}\footnote{http://heasarc.gsfc.nasa.gov/docs/swift/results/}.  Hard X-ray sources are identified in the BAT survey with position localizations of about 4\arcmin.  These positions were checked against the archives of X-ray telescopes with high spatial resolution like {\it Chandra} and {\it XMM-Newton} in order to identify an X-ray counterpart to the BAT source.  Sources with no historical observations were observed with {\it Swift}/XRT in order to identify a counterpart.

The BAT Seyfert sample is ideal for our study.  These sources are nearby, thus maximizing the likelihood of being bright at $\gamma$-rays.  The redshift distribution of the \bat Seyferts is highly biased towards $z \sim 0.03$.  Since BAT is insensitive to the effects of dust and Compton-thin obscuration, the BAT AGN sample can be considered an uniform and complete flux-limited sample of AGN in the local Universe.  The sample includes classical objects that are well-studied at other wavelengths as well as relatively less well-known sources.  Currently, there is an active program to obtain imaging and spectroscopic data at infrared, optical, and X-rays \citep[][and references therein]{winter}.  Archival VLA data are also available through NVSS \citep{nvss} and FIRST \citep{first} for most of these sources.  

\subsection{Catalog Matching}

The initial \lat Seyfert sample was selected by cross-correlating the positions from the 58-month \bat catalog \citep{batcat} with those from the 11-month \lat Point Source Catalog (PSC).  The PSC dictates that the sources must be detected with a minimum signal-to-noise of 4-$\sigma$.  The criterion for a positional match was set to be within 10 arcminutes.  Of the 1099 \bat sources, 81 objects were matches between the \bat and \lat catalogs.  After the exclusion of obvious non-AGN objects such as neutron stars, pulsars, local starbursts, objects with unknown redshifts, and those without prior identifications, we reduced the total number of objects to 53.  Of these, a majority are known blazars.  We further reduced our sample to a total of five objects that are not blazar candidates \citep[e.g.,][]{cgrabs, blazars}.  Only two of these sources have optical counterparts within the \lat error circle.  These are NGC~1068 and NGC~4945.  

Both \lat and \bat have large fields of view and the errors associated with source positions are therefore large.  We performed Monte Carlo simulations to determine the rate of false matches between the two catalogs.  We randomized the \bat catalog positions by adding or subtracting a random number greater than 10 arcminutes (our positional match criterion) to the right ascension and declination from the \bat data.  The randomized \bat positions are then matched to the \lat PSC positions.  Assuming no contamination from sources in the Galactic disk, the average random matches between the two catalogs is 3.3 from 100 simulations.  This implies that there is $\sim$4\% (3.3 out of 81 matches) false match rate between the two catalogs.

\subsection{Data Reduction}

To potentially extend the number of {\it Fermi} detections, we analyzed $\sim$2.1 years of \lat data in the 1--100~GeV energy range for our sample, spanning from August 4, 2008 to September 2, 2010.  The data reduction was performed using the {\it Fermi Science Tools} version v9r15p2.  We followed the \lat data analysis threads provided by the Fermi Science Support Center (FSSC)\footnote{http://fermi.gsfc.nasa.gov/ssc/data/analysis/scitools/.}.  In particular, we used the P6V3 instrument response.  Using the more extensive 2-year data, we determined the detection significance of the objects based on binned likelihood analysis.  Again, only two objects, NGC~1068 and NCG~4945, are detected by \lat with significant signal to noise ratios and have optical counterparts within the positional error circle.  The general properties of these two sources and modeling statistics are listed in Table~\ref{tab:sample}.

\begin{figure}
\epsscale{.8}
\centering
\plotone{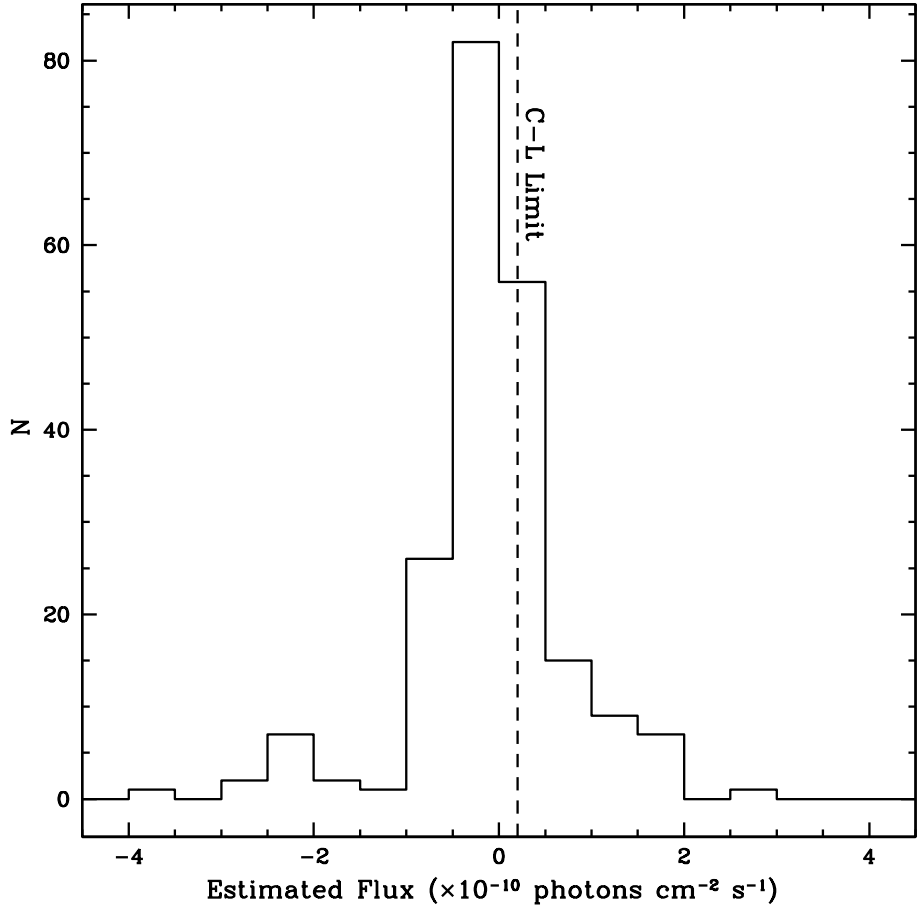}
\caption{Distribution of source fluxes as derived from individual source analysis.  These fluxes were derived by assuming point sources at the location of the Seyfert galaxies and a single power law model with $\Gamma=2.4$.  The vertical dashed line represents the limit from Composite Likelihood Analysis of 215 BAT Seyferts ($\sim 2 \times 10^{-11}$ photons~cm$^{-2}$~s$^{-1}$). }
\label{fig:ulhist}
\end{figure}

While 519 Seyferts were identified in the 58-month catalog, 28 were classified as such based on X-ray spectra and not the traditional optical methods.  Therefore, only 491 Seyferts from the \bat catalog were considered in our analysis due to uncertain classifications.  For fields centered on 276 of the 491 Seyferts, we find that the likelihood analysis is unable to successfully fit all of the $\gamma$-ray photons.  These fields are associated with very luminous $\gamma$-ray objects (e.g., 3C~273 and 3C~279) and were eliminated from subsequent analysis because the contaminating flux at the target position introduced unacceptably large uncertainties in the target flux estimates.  For the remaining 215 fields, we measured the fluxes of the undetected sources using the binned likelihood analysis.  Given the position of the \bat source, we assumed a $\gamma$-ray photon index of 2.4 for the input power law model to estimate the flux for a point source as if the galaxy were detected.  The value of 2.4 was assumed because this is the spectral index of the \lat background after the removal of the resolved sources \citep{fermiegb}; if the Seyferts make up the unresolved background, their individual spectral shapes must approximate the shape of the total spectrum.  The assumed spectral index is also consistent with the measured spectral indices of NGC~1068 and NGC~4945 (Table~\ref{tab:sample}).  Figure~\ref{fig:ulhist} shows the distribution of these derived fluxes. The distribution of the measured fluxes appear to follow a narrow Gaussian distribution centered at zero with $\sigma$ of $\sim 1 \times 10^{-10}$~photons~cm$^{-2}$~s$^{-1}$.  The negative values are due to variations in the background from both instrumental noise and the accuracy of the background model.

\subsection{Stacking Analysis}

Following the individual source analysis, we performed a stacking analysis for objects with non-detections.  This allows us to determine an upper limit to the $\gamma$-ray flux of the \bat Seyferts.  An updated version (v9r18p6) of the {\it Fermi Science Tools} released in November 2010 was used for this analysis.  We modified the Python script for composite likelihood to apply an assumed model to all of the fields simultaneously with a single model.  The script starts with the ``best-fit'' model of each field centered on a BAT source from the individual analysis and an inferred model for the Seyfert of interest.  This upper limit model is linked to be the same for all targets.  For this, we assume a power law with a fixed $\Gamma = 2.4$ and only allow the flux (normalization) value to vary.  The upper limit is determined when the $\Delta$TS value from the composite likelihood is 2.706 for one degree of freedom (at the 90\% confidence level).  

The stacking analysis estimates the upper limit to be $\sim 2 \times 10^{-11}$ photons~cm$^{-2}$~s$^{-1}$ in the {\it Fermi} band, approximately five times lower than the 1$-\sigma$ of the distribution from individual analysis (Figure~\ref{fig:ulhist}).  The stacking upper limit corresponds to an energy flux of $\sim 9 \times 10^{-14}$~ergs~cm$^{-2}$~s$^{-1}$.  Assuming the median redshift of the 215 stacked objects ($z \sim 0.031$), this implies a \lat luminosity limit of $\sim 3 \times 10^{41}$~ergs~s$^{-1}$.  This is approximately 3 and 18 times the \lat luminosities of NGC~1068 and NGC~4945, respectively, the nearest known Seyferts in the \bat sample (Table~\ref{tab:sample}).  Thus, the lack of detection of more distant Seyferts is likely a sensitivity issue.  

\subsection{New \lat Sources}

From the individual fits, we have identified 120 new extragalactic \lat sources with significant detections ($\sigma > 4$) in the fields of the \bat sample (Table~\ref{tab:nonbat}).  These sources were not included in the PSC.  The optical counterparts of these objects were identified as objects within 10 arcminutes of the nominal \lat positions.  When objects have multiple counterparts within the error circle, the one with a radio counterpart and the brightest is selected.  A majority of these 120 new sources have optical counterparts that are previously identified as quasars, flat-spectrum radio quasars (FSRQs), or BL Lac objects.  Only 17 of these have \bat counterparts; all but one has \bat detection threshold below $\sigma < 4$ (a selection criterion for the 58-month catalog).  The fact that these blazars were undetected by \bat is consistent with luminous blazars being brighter in $\gamma$-ray than hard X-ray \citep[e.g.,][]{sam10}.


\section{Discussion}
\label{sec:disc}

\subsection{Starburst Contribution}

While the current article aims to address the contribution of radio quiet AGNs to the EGB, it should be noted that starbursts may also contribute to the $\gamma$-ray output.  \citet{lenain} suggested that a starburst contributes significantly to the $\gamma$-ray luminosity of NGC~4945.  From their analysis, NGC~4945 falls on a linear relationship between the supernova rate, the total gas mass, and the $\gamma$-ray luminosity along with M~82, NGC~253, the Large Magellanic Cloud, and the Milky Way.  However, the higher detection rate of starbursts as compared to Seyferts by \lat does not necessarily imply starbursts are bigger contributors to the EGB than active nuclei.  Given the high dependence of detection on the intrinsic $\gamma$-ray luminosity and the distance of the objects, radio quiet AGNs may simply be harder to detect because there are more nearby starbursts than Seyferts.  
We can set an upper limit on the detectability of a pure starburst galaxy at the median distance of the \bat Seyferts by scaling the $\gamma$-ray luminosity upper limit by that of M~82 ($\sim 4 \times 10^{40}$~erg~s$^{-1}$).  This indicates that the star formation rate cannot be more than about 7 times what is currently seen in M~82, or else it would already have been detected.  The corresponding upper limit on the star formation rate necessary to produce this level of $\gamma$-ray emission is 45~M$_\odot$ per year.  This star formation rate is far higher than the typical range seen in Seyfert galaxies \citep[SINGS;][]{sings}.  Thus, it is unlikely for star formation to be the only, or even major, source of $\gamma$-ray emission in these galaxies.


\subsection{Radio Loudness}

In the third scenario for $\gamma$-ray production outlined in \S~\ref{sec:intro} where the jets from the AGN interact with its local environment, these sources are expected to be radio-loud.  The classical way of measuring radio loudness is that of \citet{kellerman}, by the radio-to-optical luminosity ratio ($R_o$).  By their definition, radio loud objects have radio luminosities at 5 GHz ten times that of their B-band luminosities, or  $R_o > 10$.  However, \citet{rx} noted that optical observations may be subject to obscuration and thus $R_o$ may be over estimated for some objects.  For many galaxies, the large optical apertures may also include contributions from stellar light, thus underestimating $R_o$.  Since X-ray (2--10~keV) observations are less likely to be affected by dust, \citet{rx} defined $R_X$, the radio-to-2--10~keV luminosity ratio,  as their measure of radio loudness.  Based on their study, radio loud objects have $\log R_X > -4.5$.  This is consistent with the $R_X$ boundary ($\log R_X > -4.3$) established by \citet{lafranca} using data from 1600 AGNs.

For objects with very high column densities ($N_H \sim 10^{24}$~cm$^{-2}$), even the 2--10~keV luminosity can be suppressed.  Thus, $R_X$ for Compton-thick, or nearly Compton-thick, sources may be inaccurate.  At above 10~keV, absorption is less likely to affect the X-ray measurements.  We therefore use the 14--195~keV \bat measurements to define $R_{X, BAT} = L_{1.4 GHz}/L_{X,BAT}$ and to determine radio loudness.  To establish the boundary between radio-loud and radio-quiet objects in $R_{X, BAT}$, we first determined the distribution of $R_{X, BAT}$ for Seyfert 1.0s.  From the \bat sample, there are 169 galaxies classified as Seyfert 1.0s. These objects are the least likely to have obscuration in the line of sight.  Of these, 31 have high resolution 1.4~GHz measurements from the FIRST survey.  The FIRST images confirm that these Seyfert~1.0 sources do not have any extended radio emission.  The range of $\log R_{X, BAT}$ for the Seyfert~1.0s is between --6.3 and --4.7, with a median value of --5.7.  

Given the observed range in $R_{X,BAT}$ parameters, what is the cut-off for radio-loudness?  The \citet{lafranca} results suggest the $\log R_X$~--4.3 boundary.  Of the 31 Seyfert~1.0s with FIRST fluxes, seven have published 2--10~keV fluxes.  For unabsorbed sources such as these, a direct relationship between the \bat and 2--10~keV fluxes is expected.  A linear regression analysis comparing the \bat and 2--10~keV fluxes for all seven sources results in a correlation coefficient of $R^2 = 0.09$, suggesting no correlation even for these unobscured sources.  However, after the removal of a single outlier (NGC~985, a ring galaxy), the correlation improves to $R^2 = 0.95$, consistent with expectation albeit with a small number of objects.  The data imply
\begin{equation}
f_{2-10~keV} = 0.42 f_{BAT}.
\end{equation}
This relationship is approximately that assumed by \citet{rigby}, who used the conversion $f_{2-10~keV} = 0.37f_{BAT}$ derived from AGN spectral templates constructed by \citet{marconi}. Using our empirical relation, the $\log R_{X, BAT}$ values for the 31 Seyfert~1.0s correspond to $\log R_{X}$ where $\log R_X \sim \log R_{X, BAT} + 0.4$.  The empirical relation is approximately consistent with a power law model with a photon index of 1.7.  Therefore, using the \citet{lafranca} cut-off, we define an object to be radio loud if
\begin{equation}
\log \frac{f_{1.4~GHz}}{f_{BAT}} > -4.7.
\end{equation}
The advantage of using \rbat as the radio loudness selection over \rx is clearly demonstrated in NGC~4945.  A well-known Compton-thick object, its $\log R_{X}$ value (--3.6) places it well into the radio loud category whereas its $\log R_{X, BAT}$ value (--4.3) puts it near the cutoff.  

\begin{figure}
\epsscale{.8}
\centering
\plotone{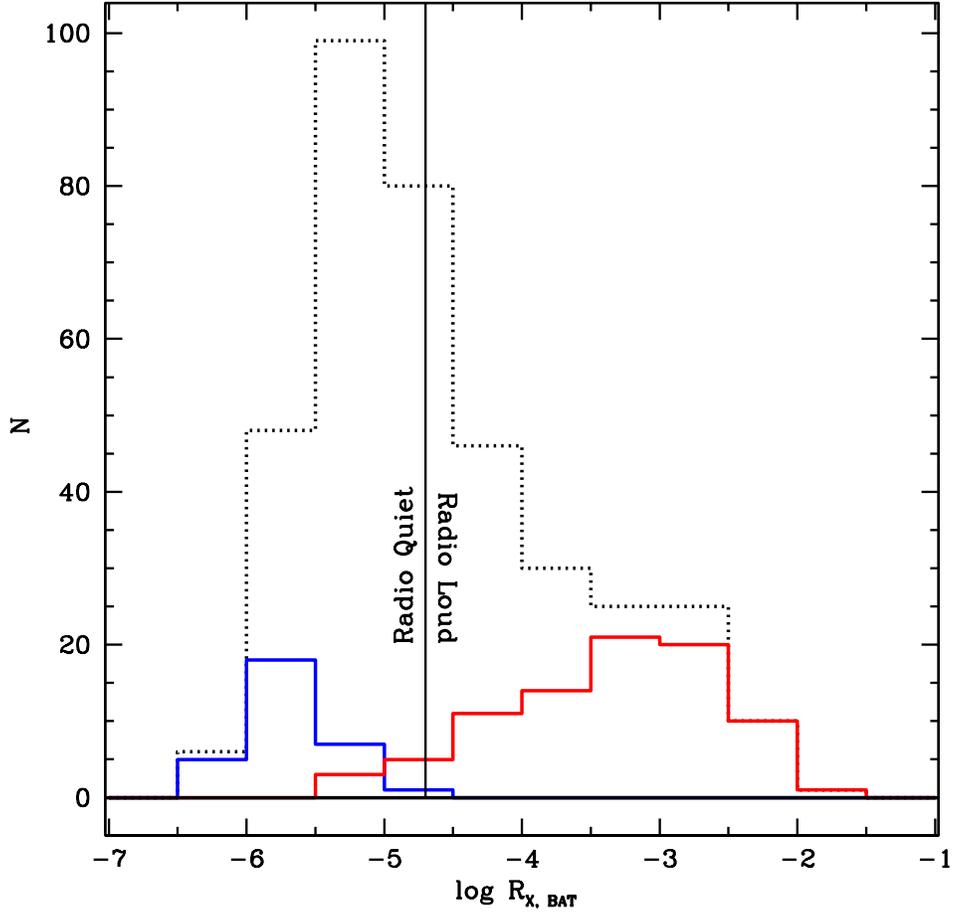}
\caption{Distribution of radio-loudness ($\log R_{X, BAT}$) amongst the extragalactic 58-month \bat sources.  The dotted histogram represents all of the extragalactic sources in the \bat sample that have available 1.4~GHz fluxes from FIRST, NVSS, or reliable VLA measurements from the literature (i.e. the parent sample).  Although the distribution appears to be skewed toward radio quiet sources, this is uncertain because it depends on the availability of radio data.  The blue histogram represents the sub-sample of radio-quiet Seyfert~1s and the red represents that of the sub-sample of blazars and FSRQs identified in the \bat sample.  There is a clear distinction between the two classes and that our radio-loudness cutoff is able to separate between the two classes.  NGC~1068 and NGC~4945 are very similar to the blazars in terms of radio loudness with $\log R_{X, BAT}$ of --3.7 and --4.3, respectively.  Nearly all of the radio loud objects have already been identified by {\it Fermi}.}
\label{fig:rbathist}
\end{figure}

Figure~\ref{fig:rbathist} shows the distribution of $\log R_{X, BAT}$ for the extragalactic \bat sources from the 58-month catalog.  The distribution is skewed toward the radio quiet sources, but it is not necessarily an accurate representation of the complete sample because it depends on the availability of reliable radio data.  The two {\it Fermi}-detected sources appears to be two of the most radio-loud objects in the \bat sample, very similar to the $\log R_{X, BAT}$ range of blazars and FSRQs detected by {\it Fermi}/LAT.  Nearly all of the radio loud objects from the \bat catalog have already been detected by {\it Fermi}.  




\subsection{Implications for the EGB}

Blazars (BL Lac objects and FSRQs) are known to contribute to about 16\% of the EGB at above 100~MeV\citep{fermiegb}\footnote{The estimate of 16\% contribution from blazars given in \citet{fermiegb} is the fraction relative to the unresolved \lat background rather than the total EGB.  \citet{ghirlanda} and \citet{inoue} estimate that blazars contribute to $\sim$45\% of the total EGB.}; the origin of the remaining fraction is still a mystery.  Radio quiet AGNs like Seyfert galaxies, though intrinsically faint, may turn out to be a significant source of the EGB if there is a large number of Seyferts.  Using \bat data, \citet{ajello} concluded that blazars, specifically FSRQs, begin to dominate the cosmic X-ray background (CXB) above a few hundred keV.  Seyfert galaxies dominate the CXB at below this energy and in the \bat energy band \citep{gilli}.  Thus, a relatively high $\gamma$-to-X-ray flux ratio may imply radio quiet AGNs are a significant source of the EGB (assuming no cosmic evolution).

\begin{figure}
\epsscale{.8}
\centering
\plotone{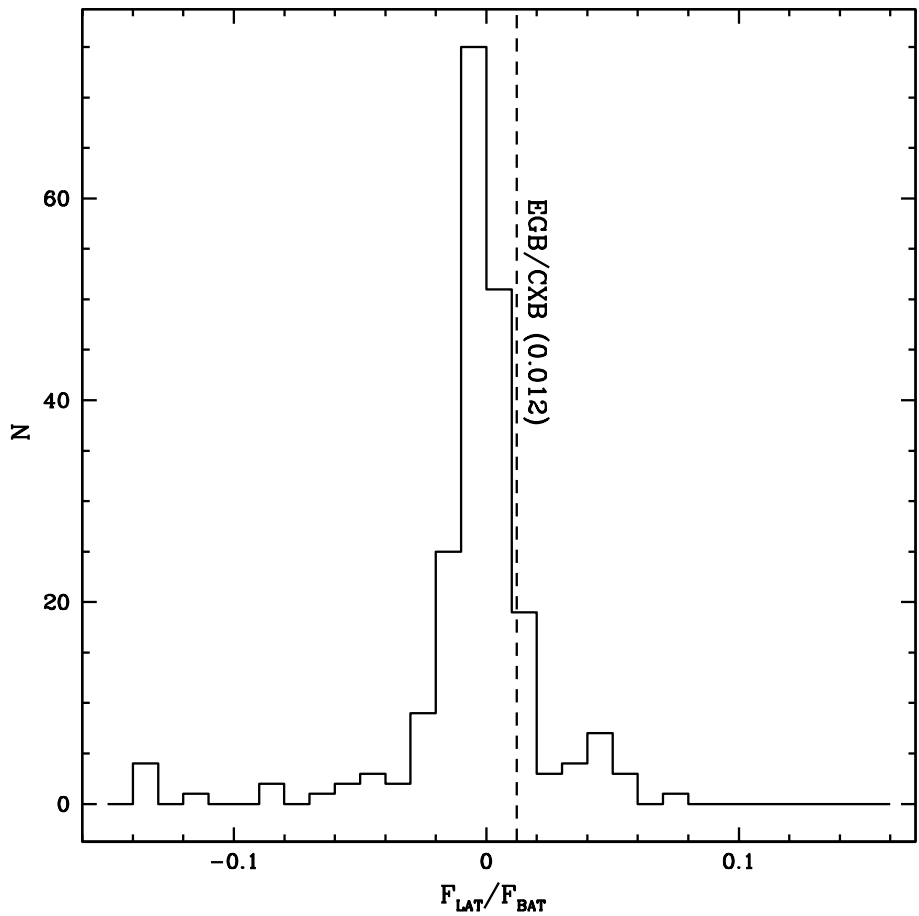}
\caption{Distribution of the upper limits to the $\gamma$-to-X-ray flux ratio from individual source analysis.  The $\gamma$-ray fluxes are those from Figure~\ref{fig:ulhist}.  Assuming no evolution as a function of redshift, the majority of the distribution lies below the EGB/CXB ratio (1.2\%).  This implies that radio quiet Seyferts are not a source of the EGB.  The flux ratio also suggest that radiatively inefficient accretion is not a $\gamma$-ray production mechanism in Seyfert galaxies.}
\label{fig:xgrathist}
\end{figure}

Figure~\ref{fig:xgrathist} shows the distribution of the upper limits of the $\gamma$-to-X-ray flux ratio of the {\it Fermi}-undetected sources.  The EGB/CXB ratio as derived from {\it Fermi} 1--100~GeV data \citep{egbcalc} and {\it Swift} data \citep{cxbcalc} is 1.2\% which is above the limits placed on the $\gamma$-to-X-ray flux ratios by \lat.  Therefore, the radio quiet Seyferts are not a significant source of the EGB.  The $\gamma$-to-X-ray flux ratio distribution also suggests that inefficient accretion flow around a black hole with a high spin value is not a viable mechanism for $\gamma$-ray emission (see \S~\ref{sec:intro}) as the model suggests an observed $\gamma$-to-X-ray flux ratio of unity.

\begin{figure}
\epsscale{.8}
\centering
\plotone{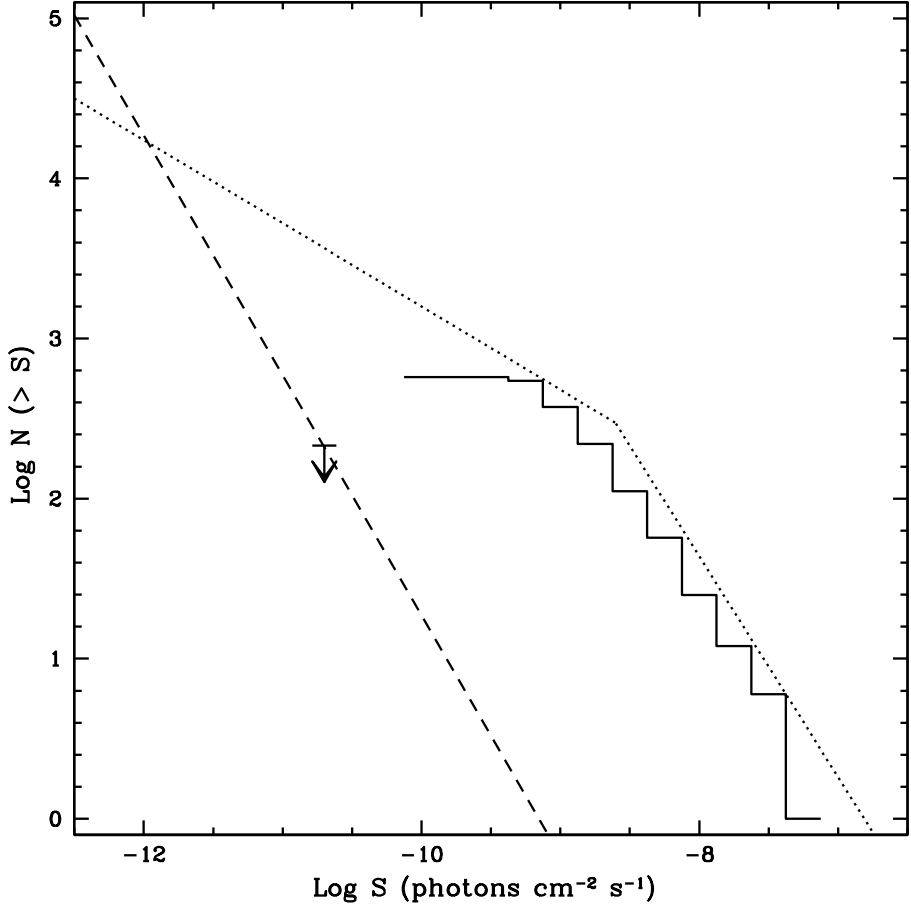}
\caption{Cumulative $\log N - \log S$ of {\it Fermi} sources.  The histogram represents the distribution of 573 detected objects associated with blazars (BL Lac sources + FSRQs) in the {\it Fermi} 1-year Catalog.  The dotted line is the empirical $\log N - \log S$ relation derived for these sources in the 1--100~GeV band from \citet{fermiegb}.  There is a break in the power law relation at $2 \times 10^{-9}$~photons~cm$^{-2}$~s$^{-1}$ with a differential power law slope of 2.4 at fluxes above the break and 1.5 below.  Also plotted is the upper limit derived from 2.1 years of {\it Fermi} data on the \bat Seyferts.  The dashed line has a differential power law slope of 2.5, assuming the same slope as that found by {\it Swift} \citep{ajello} and scaled to intersect the derived upper limit from {\it Fermi} data (labeled with the arrow).   Contributions by Seyfert galaxies would not begin to dominate the EGB until below $\sim 10^{-12}$~photons~cm$^{-2}$~s$^{-1}$, far below the current sensitivity of {\it Fermi} data.  In order for the integrated flux from the Seyfert galaxies to equal that of the ``missing'' EGB fraction, the instrument needs to have point source sensitivity $\lesssim$10$^{-23}$~photons~cm$^{-2}$~s$^{-1}$.}
\label{fig:logns}
\end{figure}

The cumulative $\log N - \log S$ in Figure~\ref{fig:logns} suggests that the \bat Seyferts would only begin to dominate the EGB over the blazars at very low flux levels ($\lesssim 10^{-12}$~photons~cm$^{-2}$~s$^{-1}$) if we assume Seyferts in fact produce $\gamma$-rays.  This limit is far below the current sensitivity of {\it Fermi} data (about ten times below our upper limit).  In the 1--100~GeV band, blazars actually only contribute to $\sim$16.6\% of the EGB intensity \citep{fermiegb}.  If we assume that Seyferts make up the rest of the EGB intensity and follow the $\log N -\log S$ relation derived from \bat data, then it would require the ability to detect individual sources at the $\lesssim$10$^{-23}$~photons~cm$^{-2}$~s$^{-1}$ level for the integrated flux to equal that of the ``missing'' $\gamma$-ray background, far below the source confusion level of {\it Fermi} at 1~GeV.  The contribution of the Seyferts are so small that it would necessitate the detection of a large number of these faint sources to make up the background.   Conversely, at the confusion limit of \lat at 1~GeV, the limiting flux is $\lesssim$1$\times 10^{-12}$~photons~cm$^{-2}$~s$^{-1}$, assuming the same $\log N - \log S$ slope.  As {\it Fermi} continues to scan the $\gamma$-ray sky, it may be possible to identify $\gamma$-ray emission from radio quiet AGNs near the end of the nominal lifetime of {\it Fermi} ($\sim$10 years).

\section{Summary}
\label{sec:sum}

From our analysis of 2.1 years of \lat data on 491 Seyfert galaxies selected from the 58-month \bat catalog, we derived upper limits to the $\gamma$-ray flux and luminosity of radio quiet AGNs.  
We defined a new radio loudness parameter ($\log R_{X, BAT}$) which confirms that only radio loud objects have been isolated and identified by {\it Fermi}.  The cumulative $\log N - \log S$ of the \bat Seyferts suggests that radio quiet AGNs would only begin to dominate the EGB over blazars at a flux level of $\lesssim 10^{-12}$~photons~cm$^{-2}$~s$^{-1}$.  

\acknowledgements

We are grateful to the referee for providing insightful comments that improved the manuscript.  We thank Wayne Baumgartner and the {\it Swift}/BAT team for providing the BAT 58 month catalog ahead of its release.  We made use of the NASA/IPAC Extragalactic Databased (NED), which is operated by the Jet Propulsion Laboratory, Caltech, under contract with NASA.  We acknowledge support by NASA through the \textit{Fermi} General Observer Program.


\begin{deluxetable}{llllccccc}
\tablecolumns{9}
\tabletypesize{\scriptsize}
\tablecaption{The Sample}
\tablewidth{0pt}
\tablehead{
\colhead{\lat ID} & \colhead{Association} & \colhead{$z$} & \colhead{Type} &\colhead{S/N}&\colhead{D$_{\rm L}$}&\colhead{$\Gamma$}&\colhead{L$_\gamma$}&\colhead{$\log R_{X, BAT}$}\\
\colhead{(1)} & \colhead{(2)} & \colhead{(3)} & \colhead{(4)}& \colhead{(5)} & \colhead{(6)}&\colhead{(7)}&\colhead{(8)}&\colhead{(9)}
}
\startdata
                 1FGL J0242.7+0007&NGC 1068& 0.004&  Sy 2&9.1&12.6&2.35$\pm$0.12&$1.0 \times 10^{41}$&--3.7\\
                  1FGL J1305.4--4928 & NGC 4945  & 0.002  &  Sy 2&11.4&11.1&2.32$\pm$0.08&$1.7 \times 10^{40}$ &--4.3\\
\enddata

\tablecomments{
Col.(1): \lat source identifier.  Col.(2): Source association.  Col.(3): Redshift.  Col.(4): Optical spectral type, from NED.  Col.(5): Approximate detection significance ($\sigma$) in the 100~MeV -- 300~GeV range from the 2-year \lat data.  Col.(6): Luminosity distance in Mpc derived relative to the reference frame defined by the 3K Microwave Radiation Background (e.g., NED).  Col.(7): Photon index in the 1--100~GeV energy range as derived from \lat data.  Col.(8): $\gamma$-ray luminosity in units of ergs~cm$^{-2}$~s$^{-1}$ in the 1--100~GeV energy range.  Col.(9): $\log R_{X, BAT}$.
}
\label{tab:sample}
\end{deluxetable}

\begin{deluxetable}{ccclcl}
\tablecolumns{6}
\tabletypesize{\scriptsize}
\tablecaption{Non-BAT Sources}
\tablewidth{0pt}
\tablehead{
\colhead{RA}&\colhead{Dec}&\colhead{Sign.}&\colhead{OptID}&\colhead{BAT?}&\colhead{Comments}\\
\colhead{(1)} & \colhead{(2)} & \colhead{(3)} & \colhead{(4)}& \colhead{(5)} & \colhead{(6)}
}
\startdata	
11.3971	&	--37.2896	&	7.9	&	[VCV96] Q0042--3734 (5$\farcm$7)	&	\nodata	&	Quasar at z=2.1	\\
11.4565	&	12.1716	&	7.2	&	RX J0045.6+1217 (7$\farcm$1)	&	2.0	&	X-ray source with radio and IR counterparts	\\
11.6186	&	11.9891	&	4.5	&	1RXS J004700.8+115827 (8$\farcm$0)	&	\nodata	&	X-ray source	\\
14.0353	&	--21.2382	&	7.0	&	2MASX J00561157--2109221 (3$\farcm$0)	&	\nodata	&	Unknown galaxy type\\
17.5903	&	61.4336	&	19.4	&	NVSS J010953+612231 (4$\farcm$8) 	&	\nodata	&	Radio source	\\
21.2783	&	--22.8700	&	6.6	&	GALEX 2674762193050796919 (4$\farcm$9)	&	\nodata	&	Possible quasar \\
21.4997	&	--25.9199	&	4.4	&	2dFGRS S148Z245 (2$\farcm$6)	&	\nodata	&	Galaxy of unknown type at z=0.11\\
22.0088	&	--8.2834	&	4.7	&	MCG--02--04--055 (6$\farcm$8)	&	\nodata	&	Galaxy of unknown type at z=0.023 	\\
28.3368	&	1.5573	&	7.0	&	2MASX J01525339+0127497 (7$\farcm$7)	&	3.0	&	IR source and X-ray source	\\
31.6304	&	--11.8300	&	9.7	&	CGRaBS J0206--1150 (1$\farcm$5)	&	\nodata	&	FSRQ	\\
34.5728	&	18.3418	&	5.4	&	IRAS F02150+1808 (7$\farcm$2)	&	\nodata	&	IR source	\\
34.8650	&	36.5856	&	7.8	&	V Zw 217 (0$\farcm$6)     	&	\nodata	&	Galaxy of unknown type	\\
39.0939	&	--61.9344	&	4.0	&	2dFGRS S913Z520 (1$\farcm$4)	&	\nodata	&	Galaxy of unknown type\\
43.8035	&	32.1962	&	4.4	&	B2 0252+32 (9$\farcm$4)	&	\nodata	&	Radio source \\
47.3045	&	10.5395	&	12.5	&	PKS J0309+1029 (3$\farcm$9)	&	\nodata	&	FSRQ\\
50.6060	&	--37.4430	&	5.2	&	ESO 301--G 008 (2$\farcm$7)   	&\nodata	&	Starburst, a dS0 in cluster	\\
51.8087	&	--15.4318	&	4.7	&	PMN J0327--1529 (4$\farcm$7)	&	\nodata	&	Radio source	\\
52.9615	&	--61.7174	&	4.4	&	IRAS F03305--6158 (5$\farcm$8)	&	\nodata	&	IR source	\\
53.2862	&	30.9969	&	8.6	&	87GB 033012.2+304428 (5$\farcm$3) 	&	\nodata	&	Radio source; near Galactic star-forming region	\\
53.7967	&	--13.9383	&	4.4	&	NVSS J033512--135703 (0$\farcm$8)	&	\nodata	&	Radio source	\\
54.7493	&	--12.5188	&	4.2	&	GALEX 2692741389941212225 (3$\farcm$4)	&	2.2	&	Possible quasar \\
55.0913	&	--21.3934	&	5.6	&	PKS 0338--214 (5$\farcm$2) 	&	\nodata	&	FSRQ	\\
59.3980	&	6.4552	&	6.4	&	NVSS J035702+063015 (8$\farcm$7)	&	\nodata	&	Radio source 	\\
66.1765	&	--53.4321	&	5.3	&	CRATES J0425--5331 (6$\farcm$9)	&	\nodata	&	FSRQ\\
67.7599	&	--60.3547	&	10.1	&	ESO 118--G 029  (3$\farcm$3)	&	\nodata	&	S0 galaxy	\\
69.6320	&	--45.3140	&	10.7	&	PKS 0437--454 (6$\farcm$2) 	&\nodata	&	Blazar candidate	\\
72.4882	&	11.6054	&	5.5	&	NVSS J044929+113216 (7$\farcm$8)	&	\nodata	&	Radio source	\\
80.4948	&	35.7555	&	5.6	&	NVSS J052202+355229 (6$\farcm$9)	&	\nodata	&	Radio source \\
80.6114	&	32.9544	&	5.8	&	[KLK2001] 940930.26 (6$\farcm$2)	&	\nodata	&	GRB 	\\
81.3636	&	--60.2304	&	5.4	&	SUMSS J052542--601341 (1$\farcm$8)	&	\nodata	&	Radio source 	\\
84.9618	&	--54.2098	&	9.9	&	PKS 0539--543 (9$\farcm$9)  	&	\nodata	&	FSRQ	\\
109.2407	&	45.4418	&	5.1	&	NVSS J071708+452203 (4$\farcm$7)	&	\nodata	&	Radio source\\
113.6889	&	50.3033	&	7.7	&	CGRaBS J0733+5022 (9$\farcm$3)  	& \nodata	&	FSRQ 	\\
117.0590	&	--16.7764	&	5.5	&	PMN J0748--1639  (7$\farcm$2)	&	\nodata	&	Radio source	\\
117.0859	&	45.1622	&	4.6	&	B3 0745+452 (3$\farcm$8)	&	3.2	&	Galaxy of unknown type	\\
117.2098	&	79.0598	&	6.3	&	NVSS J075043+790917 (7$\farcm$8)	&	\nodata	&	FSRQ 	\\
118.7230	&	48.4842	&	9.1	&	NVSS J075445+482350 (8$\farcm$0)	&	2.4	&	BL Lac 	\\
119.4554	&	37.7479	&	4.5	&	SDSS J075751.73+374554.3 (1$\farcm$1)	&	\nodata	&	Quasar	\\
125.8230	&	40.6476	&	8.0	&	B3 0819+408 (4$\farcm$8)	&	\nodata	&	FSRQ	\\
134.2846	&	72.0019	&	5.5	&	NVSS J085545+720543 (8$\farcm$5) 	&	\nodata	&	Radio source	\\
134.4593	&	--20.0023	&	6.0	&	IRAS  08559--1956 (9$\farcm$1)  	&	\nodata	&	IR source	\\
134.9916	&	67.4699	&	5.0	&	GALEX 2682537904855582223 (8$\farcm$1)	&	\nodata	&	Possible quasar\\
135.0210	&	--44.5387	&	6.3	&	2MASX J08595620--4433525 (2$\farcm$2)	&	\nodata	&	IR source	\\
135.4634	&	67.6414	&	5.7	&	NVSS J090038+674223 (8$\farcm$1)	&	\nodata	&	Radio source	\\
138.1924	&	--20.9321	&	6.8	&	RX J0913--2103 (8$\farcm$1)	&	\nodata	&	BL Lac	\\
139.3315	&	38.8643	&	4.7	&	4C 38.28 (6$\farcm$5)	&	\nodata	&	FR II QSO	\\
142.1210	&	--20.6005	&	6.7	&	PKS J0927--2034 (8$\farcm$8)	&	2.7	&	FSRQ 	\\
143.2444	&	86.2417	&	6.9	&	CGRaBS J0929+8612 (3$\farcm$9)	&	\nodata	&	FSRQ	\\
144.8152	&	--17.4112	&	8.6	&	PMN J0939--1731 (7$\farcm$0)	&	\nodata	&	FSRQ	\\
145.1797	&	--28.5810	&	5.0	&	PMN J0940--2829 (10$\farcm$0)	&	\nodata	&	Radio source	\\
147.1646	&	40.6008	&	5.0	&	4C 40.24 (4$\farcm$8)&	\nodata	&	FSRQ	\\
149.1279	&	47.3611	&	6.8	&	GHO 0953+4738 (3$\farcm$7)	&	\nodata	&	Galaxy cluster 	\\
149.5143	&	--13.7279	&	5.8	&	2MASX J09573643--1341492 (6$\farcm$8)	&	\nodata	&	IR source	\\
152.1692	&	--31.4742	&	4.4	&	1RXS J100819.7--312407 (6$\farcm$2)	&	\nodata	&	X-ray source	\\
152.3001	&	--31.6883	&	6.2	&	PKS J1008--3138 (5$\farcm$1)	&	\nodata	&	FSRQ	\\
153.3678	&	23.2755	&	5.0	&	CGCG 123--025 (1$\farcm$8)	&	1.6	&	IR source 	\\
154.2732	&	56.0413	&	5.0	&	GALEX 2683382329785717300 (3$\farcm$4)	&	\nodata	&	Possible quasar \\
154.6427	&	--31.2356	&	7.2	&	PKS J1018--3123 (9$\farcm$8)	&	\nodata	& FSRQ		\\
156.1451	&	0.6456	&	5.5	&	SDSS J102437.04+003926.0 (0$\farcm$9)	&	\nodata	&	Possible quasar \\
156.7702	&	74.6228	&	7.9	&	NVSS J102739+744005 (3$\farcm$5)	&	\nodata	&	Radio/IR source	\\
157.7529	&	74.6584	&	9.0	&	NVSS J103122+744158 (2$\farcm$8)	&	2.2	&	FSRQ	\\
161.5354	&	--29.4379	&	9.0	&	TXS 1043--291B (3$\farcm$7) 	&	\nodata	&	Radio source with lobes \\
163.9115	&	69.9321	&	10.6	&	2MASX J10543042+6949207 (8$\farcm$8)	&	\nodata	&	X-ray/IR source \\
164.7596	&	2.5005	&	4.7	&	PMN J1059+0225 (4$\farcm$9) 	&	1.0	&	Radio source 	\\
166.0430	&	81.2678	&	6.2	&	3B 940318 (6$\farcm$6)	&	\nodata	&	GRB 	\\
166.7367	&	--36.8196	&	6.8	&	ESO 377--G 007 (9$\farcm$6) 	&	\nodata	&	Spiral galaxy 	\\
171.9033	&	36.3784	&	8.8	&	CGRaBS J1121+3620 (5$\farcm$0)	&	2.8	&	FSRQ	\\
175.4210	&	61.3633	&	8.3	&	EXMS B1146+608 (3$\farcm$5)	&	\nodata	&	X-ray source \\
178.2468	&	49.5137	&	10.2	&	SDSS J115250.29+493220.4 (2$\farcm$1)	&	\nodata	&	Quasar	\\
183.6555	&	13.1974	&	5.0	&	4C 49.22 (4$\farcm$1)	&	\nodata	&	FSRQ	\\
183.8466	&	13.0637	&	5.7	&	SDSS J121506.93+130558.8 (4$\farcm$5)	&	\nodata	&	BAL QSO	\\
184.1765	&	18.4032	&	6.2	&	CGCG 098--127 (1$\farcm$5)	&	\nodata	&	Elliptical galaxy	\\
184.5352	&	--10.3535	&	4.9	&	2MASX J12181358--1026487 (5$\farcm$7)	&	\nodata	&	IR source 	\\
184.6073	&	--0.5755	&	5.5	&	PKS J1217--0029 (8$\farcm$3)	&	\nodata	&	BL Lac 	\\
186.1276	&	24.4276	&	6.7	&	NVSS J122437+242452 (1$\farcm$8)	&	\nodata	&	Radio source	\\
186.2145	&	21.7290	&	4.8	&	TXS 1222+220 (2$\farcm$9)	&	\nodata	&	Radio source	\\
191.6873	&	44.3953	&	5.8	&	RBS 1154 (2$\farcm$8)	&	1.4	&	BL Lac	\\
195.3049	&	33.6781	&	5.2	&	FIRST J130115.7+334329 (2$\farcm$9)	&	\nodata	&	Radio source 	\\
196.1285	&	12.0585	&	7.9	&	 SDSS J130426.15+120245.5  (1$\farcm$4)	&	\nodata	&	Quasar \\
198.4177	&	--23.8658	&	6.1	&	2MASX J13135485--2354023 (3$\farcm$9)	&	\nodata	&	IR source	\\
203.4901	&	--38.3668	&	6.0	&	2MASX J13340073--3820519 (1$\farcm$3)	&	\nodata	&	IR source	\\
204.4920	&	--24.0832	&	4.2	&	ESO 509--84 (2$\farcm$6)	&	\nodata	&	Spiral galaxy 	\\
206.5901	&	--26.0389	&	5.6	&	2MASX J13463183--2602032	(2$\farcm$3)&	\nodata	&	IR source 	\\
207.9203	&	--29.0735	&	8.8	&	PKS J1351--2912 (8$\farcm$0)	&	\nodata	&	Blazar	\\
208.3928	&	37.5322	&	6.5	&	NVSS J135344+373227 (2$\farcm$1)	&	\nodata	&	Radio source \\
210.0310	&	--14.7102	&	6.0	&	2MASX J14000692--1444062 (1$\farcm$5)	&	0.7	&	Galaxy of unknown type 	\\
214.0887	&	5.9172	&	4.7	&	3B 940621 (4$\farcm$1)	&	\nodata	&	GRB 	\\
214.1380	&	13.3386	&	6.7	&	PKS 1413+135 (8$\farcm$3)	&	\nodata	&	Radio source 	\\
215.0146	&	--8.6040	&	6.5	&	2MASX J14201987--0837228	 (4$\farcm$2)&	\nodata	&	IR source 	\\
216.9062	&	--32.8323	&	7.9	&	2MASX J14280087--3256085 (7$\farcm$9)	&	\nodata	&	Spiral galaxy \\
218.6920	&	20.5116	&	5.1	&	2MASX J14345111+2030416 (1$\farcm$2)	&	1.7	&	IR source	\\
218.8956	&	20.3668	&	6.8	&	SDSS J143539.93+202405.4 (2$\farcm$4)	&	\nodata	&	Quasar 	\\
225.4503	&	55.8271	&	7.0	&	SDSS J150153.53+555309.7 (3$\farcm$6)	&	\nodata	&	Quasar \\
230.5842	&	43.5332	&	5.5	&	CGRaBS J1521+4336 (7$\farcm$2)	&	\nodata	&	FSRQ 	\\
241.5583	&	85.1381	&	6.3	&	NVSS J160701+850215 (6$\farcm$1)	&	6.3	&	Radio source	\\
246.3093	&	43.6982	&	4.7	&	SDSS J162458.54+434036.4  (3$\farcm$1)	&	\nodata	&	Quasar 	\\
249.7440	&	41.5111	&	5.9	&	SDSS J163855.92+412937.0 (1$\farcm$2)	&	\nodata	&	Possible quasar \\
249.8025	&	39.6816	&	14.7	&	SDSS J163851.55+393759.0 (5$\farcm$0) 	&	\nodata	&	Quasar 	\\
251.1208	&	--54.6591	&	4.1	&	 WKK 7381 (4$\farcm$7) 	&	0.3	&	Galaxy of unknown type with bright nucleus	\\
252.5771	&	8.5271	&	5.5	&	CGRaBS J1650+0824 (8$\farcm$2)	&	\nodata	&	FSRQ 	\\
254.3308	&	48.2426	&	14.4	&	4C 48.41 (7$\farcm$6)	&	3.1	&	FSRQ 	\\
254.6441	&	--1.4666	&	5.0	&	PMN J1659--0127 (6$\farcm$8)	&	\nodata	&	Radio source \\
255.2459	&	39.8811	&	4.2	&	FIRST J170108.8+395443 (2$\farcm$7)	&	0.2	&	BL Lac 	\\
259.3781	&	68.6269	&	9.0	&	VII Zw 707 (5$\farcm$8)	&	\nodata	&	IR source 	\\
\tablecomments{
Col.(1): Right ascension in degrees.  Col.(2): Declination in degrees.  Col.(3): \lat detection significance.  Col.(4): Most likely optical counterpart within 10 arcminutes of the nominal position.  The value in the parantheses represent the difference between the nominal \lat position and that of the optical counterpart.  Col.(5): Approximate detection significance of the \lat source by \bat within the \bat position error of 4 arcminutes.  Col.(6): Comments about the nature of the optical counterpart to the \lat source.
}
\enddata
\label{tab:nonbat}
\end{deluxetable}


\end{document}